\documentclass[journal]{IEEEtran}

\usepackage{blindtext}
\usepackage{graphicx}
\usepackage{blindtext, graphicx}
\usepackage{subcaption}
\usepackage{graphicx}
\ifCLASSINFOpdf
\else
  
\fi

\usepackage{amsmath,amssymb}
\usepackage{amsmath}
\usepackage{algorithm}
\usepackage[noend]{algpseudocode}
\usepackage{lipsum}
\usepackage{mathtools}
\usepackage{cuted}

\usepackage{algorithm,algpseudocode}
\makeatletter
\newcommand{\algmargin}{\the\ALG@thistlm}
\makeatother
\newlength{\whilewidth}
\algdef{SE}[parWHILE]{parWhile}{EndparWhile}[1]
  {\parbox[t]{\dimexpr\linewidth-\algmargin}{%
     \hangindent\whilewidth\strut\algorithmicwhile\ #1\ \algorithmicdo\strut}}{\algorithmicend\ \algorithmicwhile}%
\algnewcommand{\parState}[1]{\State%
  \parbox[t]{\dimexpr\linewidth-\algmargin}{\strut #1\strut}}

\begin{document}
\title{Physical Layer Security in Relay Networks with Outdated Relay Selection }

\author{Shahla~Mohsenifard,~\IEEEmembership{Member,~IEEE}% <-this % stops a space
% <-this % stops a space
}


\markboth{}%
{Shell \MakeLowercase{\textit{et al.}}: Bare Demo of IEEEtran.cls for Journals}

\maketitle

\begin{abstract}
%\boldmath
In this paper, the secrecy performance of a cooperative relay network with outdated relay selection is investigated where an eavesdropper intercepts the channels between the source and the destination. The best relay is chosen among $N$ relays based on the  opportunistic relay selection algorithm, which may not be the best relay in the time of transmission because of the outdated channel state information. We derive closed-form analytical expressions for the non-zero secrecy capacity, the secrecy outage probability, and the ergodic secrecy capacity. Finally, our theoretical analysis is validated by the numerical results, and detailed discussions and insights are given.

\end{abstract}


% \fi
\IEEEpeerreviewmaketitle

\section{Introduction}

Wireless communication networks due to their broadcast nature are prone to various attacks and misuse behavior such as being overheard and intercepted by eavesdroppers. As such, security issues of wireless communications are of great importance. Although cryptographic methods can improve the communication security, as the computational ability and the power of eavesdroppers are stronger nowadays, it is more and more likely to intercept the network successfully. As a result, physical layer security has attracted increasing attention in the literature \cite{IEEEhowto:Pandey} - \cite{IEEEhowto:Hoang}. Relaying is one of the effective techniques which has received wide attention to extend the coverage area and improve communication quality in wireless networks \cite{IEEEhowto:Shi}. Thus, investigating the physical layer security of relay networks is crucial. Opportunistic relay selection is among low complexity cooperative techniques which activate only the best relay. In selecting the best relay, a delay between relay selection instant and data transmission instant may cause an outdated channel state information because of the time-varying nature of fading channels, \cite{IEEEhowto:Yadav}, which leads to a degradation in the secrecy performance of the network. 

There have been several works dealing with secrecy issues in cooperative relay systems. In \cite{IEEEhowto:Zou} AF and DF based optimal relay selection schemes are proposed to improve the wireless security against eavesdropping, where the authors consider the exact knowledge of CSI. Authors in \cite{IEEEhowto:Huang} analyzed the secrecy outage probability of a relay network while passive eavesdroppers are intercepting the channel and partial relay selection scheme is considered to select the best relay. In \cite{IEEEhowto:Fan} secrecy performance of a cooperative network under the constraint of having outdated CSI is studied. The authors assumed that relays only to assist the transmission, and there is no direct link between the source and the destination. S. Al-Qahtani et al. comprehensively investigate on the secrecy performance of opportunistic relay selection systems over Rayleigh fading channels. The partial relay selection scheme is considered, and the secrecy is studied when there is a feedback delay \cite{IEEEhowto:Al-Qahtani}. In \cite{IEEEhowto:Wang} with the aid of joint relay and jammer selection, the physical layer security of amplify-and-forward relaying networks is increased despite the effects of channel feedback delay.
In this paper, we investigate on secrecy performance of a cooperative relay network with outdated CSI in the presence of an eavesdropper. The information transmission is assisted by $N$ relays from the source to the destination, and a direct link between the source and the destination is also considered. We study the impact of outdated relay selection by deriving the analytical expressions for the secrecy performances. 

The reminder of the paper is organized as follows. Section II introduces the system and channel models. In Section III, we derive closed-form expressions of the SOP, non-zero secrecy capacity, and ergodic secrecy capacity. In section IV we present simulation results and in section V we conclude this work.

\section{System Model}
Consider a cooperative wireless network consisting of one source, one destination, and $N$ trusted relays while an eavesdropper is passively eavesdropping channels without modifying the signals. The source and the destination communicate over one direct $S$-$D$ channel and $N$ two-hop channels using relays while all channels experience Rayleigh fading. All nodes are equipped with a single antenna. 

\begin{figure}
\centering
\includegraphics[scale = 0.55]{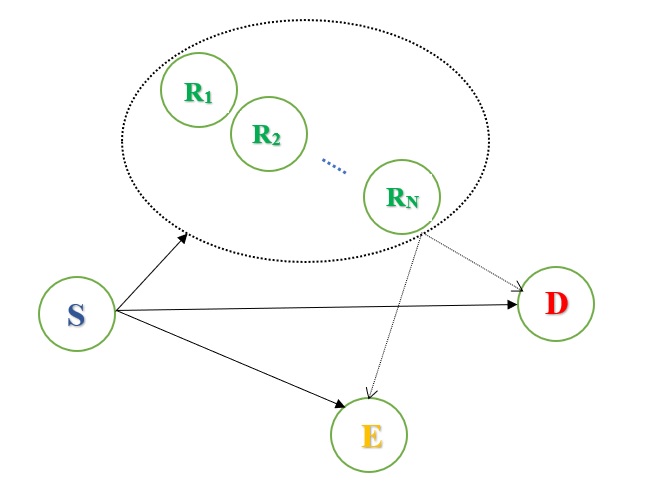} 
\caption{System model. }
\end{figure}

The thermal noise received at each authored node is modeled as a complex Gaussian random variable with zero mean and variance $\sigma_{n}^{2}$ i.e., $CN(0,\sigma_{n}^{2})$ and the thermal noise at the eavesdropper is  a complex Gaussian random variable with zero mean and variance $\sigma_{e}^{2}$ i.e., $CN(0,\sigma_{e}^{2})$ . It is assumed that the channel state information is known at the destination. The best relay is selected as $b=argmax_{n\in R}\{\gamma_{n}\}$, where $\gamma_{n}=\min\{\gamma_{SR_{n}},\gamma_{R_{n}D}\}$ is an upper bound SNR of $S-R_{n}-D$ link for relay $n$ and $R=\{1,2,...,N\}$. From \cite{IEEEhowto:T1} the total $SNR$ in the destination can be written as 
\begin{equation}
 \gamma_{opr}=\gamma_{SD}+\max_{n\in {R}}\Big(\min\{\gamma_{SR_{n}},\gamma_{R_{n}D}\}\Big)
\end{equation}
where $\gamma_{SD} ={|h_{SD}|}^2 {E_{s}}/{\sigma_{n}^2}$ is the instantaneous SNR of $S-D$ link, $\gamma_{SR_{n}} ={|h_{SR{n}}|}^2 {E_{s}}/{\sigma_{n}^2}$, and  $\gamma_{R_{n}D} ={|h_{R_{n}D}|}^2 {E_{s}}/{\sigma_{n}^2}$ are the instantaneous SNRs of $S-R_{n}$ and $R_{n}-D$ links respectively. The total SNR in eavesdropper is then 
\begin{equation}
 \gamma_{opr,e}=\gamma_{SE}+\min\{\gamma_{Sb},\gamma_{bE}\}
\end{equation}
where $\gamma_{SE}={|h_{SE}|}^2 {E_{s}}/{\sigma_{e}^2}$, $\gamma_{Sb}={|h_{Sb}|}^2 {E_{s}}/{\sigma_{e}^2}$, and  $\gamma_{bE}={|h_{bE}|}^2 {E_{s}}/{\sigma_{e}^2}$ are $S-E$, $S-b$, and $b-E$ links' SNRs respectively. From a practical point of view, some factors such as channel changes in time, feedback delays, and the channel estimation errors may cause a difference between the CSI of $S-R_{n}-D$ links required for the opportunistic relaying and the actual values.  In other words, the best relay selected according to the outdated CSI at time $t$ may not be the best relay at the time of the data transmission (time $t+\tau$). In \cite{IEEEhowto:T2} a degree of difference is considered as the power correlation coefficient between the SNR at time $t$, $\tilde{\gamma_{n}}$, and the SNR at time $t+\tau$, $\gamma_{n}$, denoted by $\rho$ where $0<\rho<1$.

\section{secrecy capacity, secrecy outage probability, and ergodic secrecy capacity}
\subsection{Non-zero secrecy capacity} 
This part characterizes the secrecy capacity of the relay network with outdated Rayleigh channels. The total capacity of the channel between the source and the destination is 
\begin{equation}
C_{M}=\frac{1}{2}\mathrm{log}_{2}(1+\gamma_{opr})
\end {equation}
and the total capacity of the channel between the source and the eavesdropper is 
\begin{equation}
C_{E}=\frac{1}{2}\mathrm{log}_{2}(1+\gamma_{opr,e})
\end {equation}

Mathematically, the instantaneous secrecy capacity can be expressed as  \cite{IEEEhowto:T2} 
\begin{equation}
C_{s}=\begin{cases} \label{eq:5}
\frac{1}{2}(\mathrm{log}_{2}(1+\gamma_{opr})-\mathrm{log}_{2}(1+\gamma_{opr,e}))  & \gamma_{opr}>\gamma_{opr,e}\\
0 &\gamma_{opr}<\gamma_{opr,e}
\end {cases}
\end {equation}

We will now consider the existence of a non-zero secrecy capacity. From (\ref{eq:5}) it follows that the secrecy capacity is positive when $\gamma_{opr}>\gamma_{opr,e}$ and is zero when $\gamma_{opr}<\gamma_{opr,e}$. Considering the independence between the channels, the probability of non-zero secrecy capacity can be written as
\begin{equation}
\begin {split}
p(C_{s}>0)&= p(\gamma_{opr}>\gamma_{opr,e}) \\
&= \int_{0}^{\infty}\int_{0}^{ \gamma_{opr}}f( \gamma_{opr}, \gamma_{opr,e})d \gamma_{opr,e}d \gamma_{opr}\\
&=\int_{0}^{\infty}\int_{0}^{ \gamma_{opr}}f( \gamma_{opr})f( \gamma_{opr,e})d \gamma_{opr,e}d \gamma_{opr}\\
\end{split}
\end{equation}

It has been shown in \cite{IEEEhowto:T2} that the probability density function (PDF) of $\gamma_{opr}$ in the network is 
\begin{equation}
\begin{split}\label{eq:7}
f_{\gamma_{opr}}(\gamma)&=\sum_{n=1}^{N}{{N}\choose {n} }\frac{(-1)^{n-1}}{\Big(\bar{\gamma}_{c}[n(1-\rho)+\rho]/n-\bar{\gamma}_{SD}\Big)} \\
 &\times\bigg[\textrm{exp}\Big(\frac{-n\gamma}{\bar{\gamma}_{c}[n(1-\rho)+\rho]}\Big)-\textrm{exp}\Big(\frac{-\gamma}{\bar{\gamma}_{SD}}\Big)\bigg]
\end{split}
\end{equation}
where ${\bar{\gamma}_{c}}=\frac{\bar{\gamma}_{SR_{n}}\bar{\gamma}_{R_{n}D}}{\bar{\gamma}_{SR_{n}}+\bar{\gamma}_{R_{n}D}}$ and $n=1,2,3,...,N$, and the PDF of  $\gamma_{opr,e}$ is
\begin{equation}\label{eq:8}
f_{\gamma_{opr,e}}(\gamma)=\frac{1}{\Big(\bar{\gamma}_{ce}-\bar{\gamma}_{SE}\Big)} \times\bigg[\textrm{exp}\Big(\frac{-\gamma}{\bar{\gamma}_{ce}}\Big)-\textrm{exp}\Big(\frac{-\gamma}{\bar{\gamma}_{SE}}\Big)\bigg]
\end{equation}
where ${\bar{\gamma}_{ce}}=\frac{\bar{\gamma}_{Sb}\bar{\gamma}_{bE}}{\bar{\gamma}_{Sb}+\bar{\gamma}_{bE}}$. So the probability of non-zero secrecy capacity can be proved as
\begin{align}
\begin{split}\label{eq:9}
p(C_s>0)=&\sum_{n=1}^{N}{{N}\choose {n} }(-1)^{n-1}\Bigg{(}1-\frac{1}{g-{\overline{\gamma }}_{SD}}\times\\
&\Bigg{[}\frac{{\overline{\gamma }}_{ce}}{{\overline{\gamma }}_{ce}-{\overline{\gamma }}_{SE}}\times \Big{(}\frac{1}{\frac{1}{g}+\frac{1}{{\overline{\gamma }}_{ce}}}-\frac{1}{\frac{1}{{\overline{\gamma }}_{SD}}+\frac{1}{{\overline{\gamma }}_{ce}}}\Big{)}-\\
&\frac{{\overline{\gamma }}_{SE}}{{\overline{\gamma }}_{ce}-{\overline{\gamma }}_{SE}}\times \Big{(}\frac{1}{\frac{1}{g}+\frac{1}{{\overline{\gamma }}_{SE}}}-\frac{1}{\frac{1}{{\overline{\gamma }}_{SD}}+\frac{1}{{\overline{\gamma }}_{SE}}}\Big{)} \Bigg{]}\Bigg{)}
\end{split}
\end{align}
where $g=\bar{\gamma}_{c}[n(1-\rho)+\rho]/n$. 

\textit{Proof}: See Appendix A for the detailed derivation.

\subsection{Secrecy outage probability}

Now we characterize the secrecy outage probability (SOP). The SOP is defined as the probability that the instantaneous secrecy capacity falls below a predefined secrecy target rate $R_{s}>0$. Thus the SOP of the system can be considered as 
\begin{equation}\label{eq:10}
p_{out}(R_{s})=p(C{s}<R_{s})
\end{equation}

It is clear that when $C_{M}<C_{E}$, $p_{out}(R_{s})=1$. When $C_{M}>C_{E}$ the SOP can be  given as
\begin{equation}
\begin{split}\label{eq:11}
p_{out}&(R_{s})=1-\sum^N_{n=1}{{N}\choose {n} }(-1)^{n-1} a \times\\
& \Bigg{(}g\ \mathrm{exp}\Big(\frac{1-2^{R_s}}{g}\Big) \bigg[\frac{1}{\frac{2^{R_s}}{g}+\frac{1}{{\overline{\gamma }}_{ce}}}-\frac{1}{\frac{2^{R_s}}{g}+\frac{1}{{\overline{\gamma }}_{SE}}}\bigg]-\\
&{\overline{\gamma }}_{SD}\ \mathrm {exp}\Big(\frac{1-2^{R_s}}{{\overline{\gamma }}_{SD}}\Big)\bigg[\frac{1}{\frac{2^{R_s}}{{\overline{\gamma }}_{SD}}+\frac{1}{{\overline{\gamma }}_{ce}}}-\frac{1}{\frac{2^{R_s}}{{\overline{\gamma }}_{SD}}+\frac{1}{{\overline{\gamma }}_{SE}}}\bigg]\Bigg{)}
\end{split}
\end{equation}
where $a=\frac{1}{{\overline{\gamma }}_{ce}-{\overline{\gamma }}_{SE}}  \frac{1}{g-{\overline{\gamma }}_{SD}}$.

\textit{Proof}: See Appendix B for the detailed derivation.

\subsection{Ergodic secrecy capacity}
The ergodic secrecy capacity serves as another important metric of wireless fading channel, which is calculated as the average of instantaneous capacity over $\gamma_{opr}$ and $\gamma_{opr,e}$. The ergodic secrecy capacity is defined as [6]

\begin{equation}\label{eq:12}
\overline{C}_{s}=\int^{\infty }_0{\int^{\infty }_0{C_sf_{{\gamma{}}_{opr}}({\gamma }_{opr})f_{{\gamma{}}_{opr,e}}({\gamma }_{opr,e}){{d{\gamma }_{opr}}}{{d{\gamma }_{opr,e}}}}}
\end{equation}

By substituting the equations (\ref{eq:5}) in (\ref{eq:12}) we have 
\begin{equation}
\begin{split}\label{eq:13}
\overline{C}_{s}=&\int^{\infty }_0\int^{{\gamma}_{opr} }_0\frac{1}{2}\mathrm{log}_{2}(1+{\gamma}_{opr})\times \\& f_{{\gamma{}}_{opr}}({\gamma }_{opr})f_{{\gamma{}}_{opr,e}}({\gamma }_{opr,e}){{d{\gamma }_{opr,e}}}{{d{\gamma }_{opr}}}\\
-&\int^{\infty }_0\int^{\infty }_{{\gamma}_{opr,e}}\frac{1}{2}\mathrm{log}_{2}(1+{\gamma}_{opr,e})\times \\&f_{{\gamma{}}_{opr}}({\gamma }_{opr})f_{{\gamma{}}_{opr,e}}({\gamma }_{opr,e}){{d{\gamma }_{opr}}}{{d{\gamma }_{opr,e}}}\\
\end{split}
\end{equation}
\begin{figure*}
\begin{align}
\begin{split}\label{eq:14}
&\overline{C_s}=\frac{1}{2\ln{2}}\frac{1}{{\overline{\gamma }}_{ce}-{\overline{\gamma }}_{SE}} \sum_{n=1}^{N}{{N}\choose {n} }\frac{{\left(-1\right)}^{n-1}}{g-{\overline{\gamma }}_{SD}}\Biggr[({\overline{\gamma }}_{ce}-{\overline{\gamma }}_{SE})\Bigg(g\ \mathrm{exp}\mathrm{}\bigg(\frac{1}{g}\bigg)\mathrm{Ei}\left(\frac{1}{g}\right)-{\overline{\gamma }}_{SD}\ {\mathrm{exp} \left(\frac{1}{{\overline{\gamma }}_{SD}}\right)\ }\mathrm{Ei}\left(\frac{1}{{\overline{\gamma }}_{SD}}\right)\Bigg)-\\
&({\overline{\gamma }}_{ce}+g)\Bigg(\frac{1}{\frac{1}{g}+\frac{1}{{\overline{\gamma }}_{ce}}} \mathrm{exp}\mathrm{}\bigg(\frac{1}{g}+\frac{1}{{\overline{\gamma }}_{ce}}\bigg)\mathrm{Ei}\left(\frac{1}{g}+\frac{1}{{\overline{\gamma }}_{ce}}\right)\Bigg)+({\overline{\gamma }}_{ce}-{\overline{\gamma }}_{SD}) \left(\frac{1}{\frac{1}{{\overline{\gamma }}_{SD}}+\frac{1}{{\overline{\gamma }}_{ce}}} {\mathrm{exp} \left(\frac{1}{{\overline{\gamma }}_{SD}}+\frac{1}{{\overline{\gamma }}_{ce}}\right)\ }\mathrm{Ei}\left(\frac{1}{{\overline{\gamma }}_{SD}}+\frac{1}{{\overline{\gamma }}_{ce}}\right)\right)+\\
&({\overline{\gamma }}_{SE}-g)\left(\frac{1}{\frac{1}{g}+\frac{1}{{\overline{\gamma }}_{SE}}}{\mathrm{exp}\left(\frac{1}{g}+\frac{1}{{\overline{\gamma }}_{SE}}\right)}\mathrm{Ei}\left(\frac{1}{g}+\frac{1}{{\overline{\gamma }}_{SE}}\right)\right)-({\overline{\gamma }}_{SE}+{\overline{\gamma }}_{SD})\left(\frac{1}{\frac{1}{{\overline{\gamma }}_{SD}}+\frac{1}{{\overline{\gamma }}_{SE}}}{\mathrm{exp} \left(\frac{1}{{\overline{\gamma }}_{SD}}+\frac{1}{{\overline{\gamma }}_{SE}}\right)\ }\mathrm{Ei}\left(\frac{1}{{\overline{\gamma }}_{SD}}+\frac{1}{{\overline{\gamma }}_{SE}}\right)\right) \Biggl]
\end{split}
\end{align}
\end{figure*}
\quad So the ergodic secrecy capacity is as in (\ref{eq:14}).

\textit{Proof}: See Appendix C for the detailed derivation.

\section{Simulation and Numerical results}

In this section, we provide numerical results to validate our analytical expressions. We consider $N=5$ relays and set the threshold secrecy rate $R_{s} = 2$ bits/s/Hz and  all links in the network experience Rayleigh fading. We assume $\bar{\gamma}_{c}=0.5 \bar{\gamma}_{SD}$, $\bar{\gamma}_{SD}={E_{s}}/{\sigma_{n}}$, $\bar{\gamma}_{ce}=0.5\bar{\gamma}_{SE}$ and  $\bar{\gamma}_{SE}={E_{s}}/{\sigma_{e}}$.
%%%%%%%%%%%%%%%%%%%%%%%%
% Fig1
\begin{figure}[H]
\centering
\includegraphics[scale =0.5]{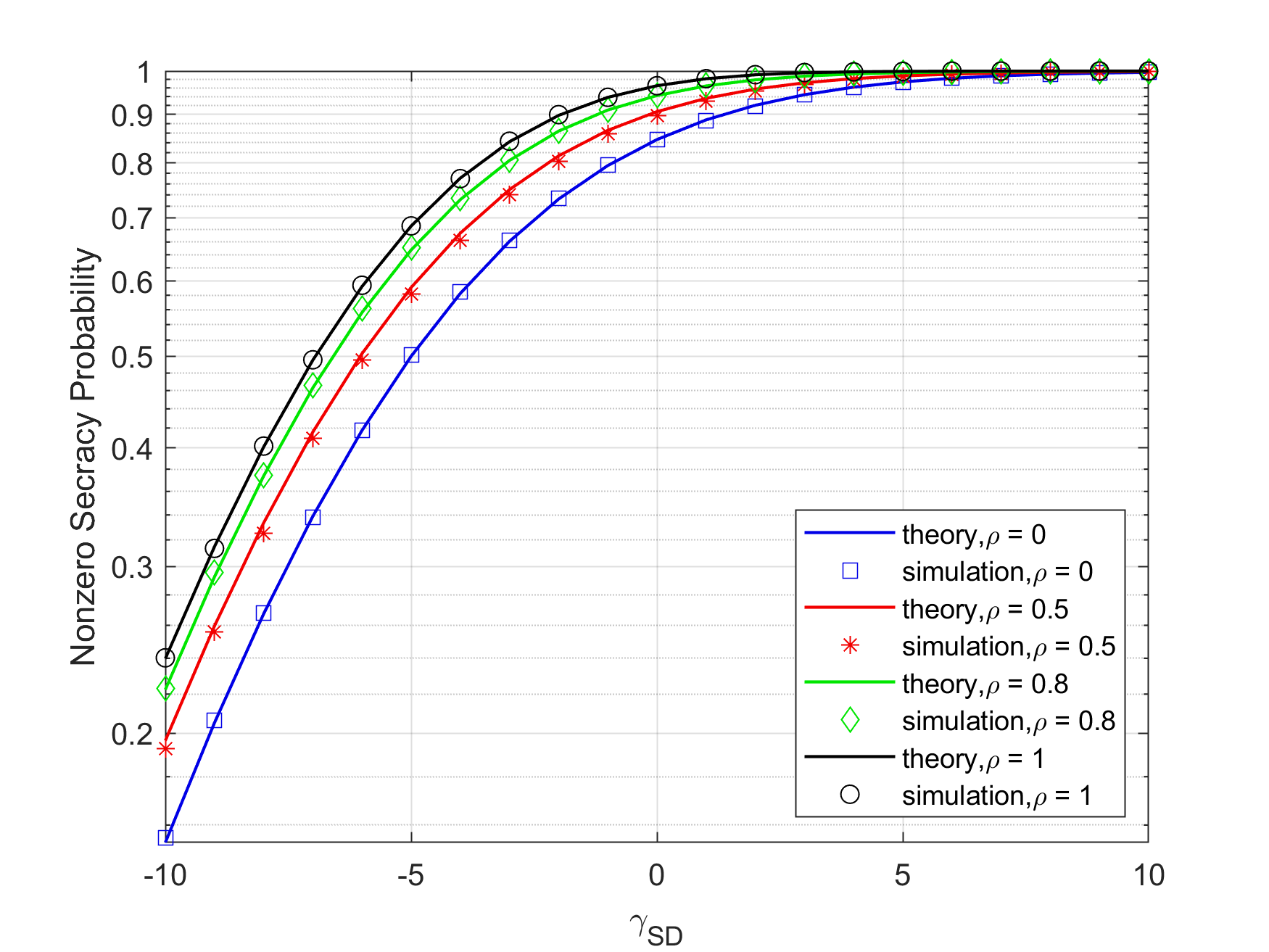} 
\caption{Non-zero secrecy probability for different values of $\rho$ and $\bar{\gamma}_{SE}=-5dB$}
  \label{fig:nonfloat}
\end{figure}
%%%%%%%%%%%%%%%%%%%%%%%%

%%%%%%%%%%%%%%%%%%%%%%%%
%Fig 2
\begin{figure}[H]
\centering
\includegraphics[scale = 0.5]{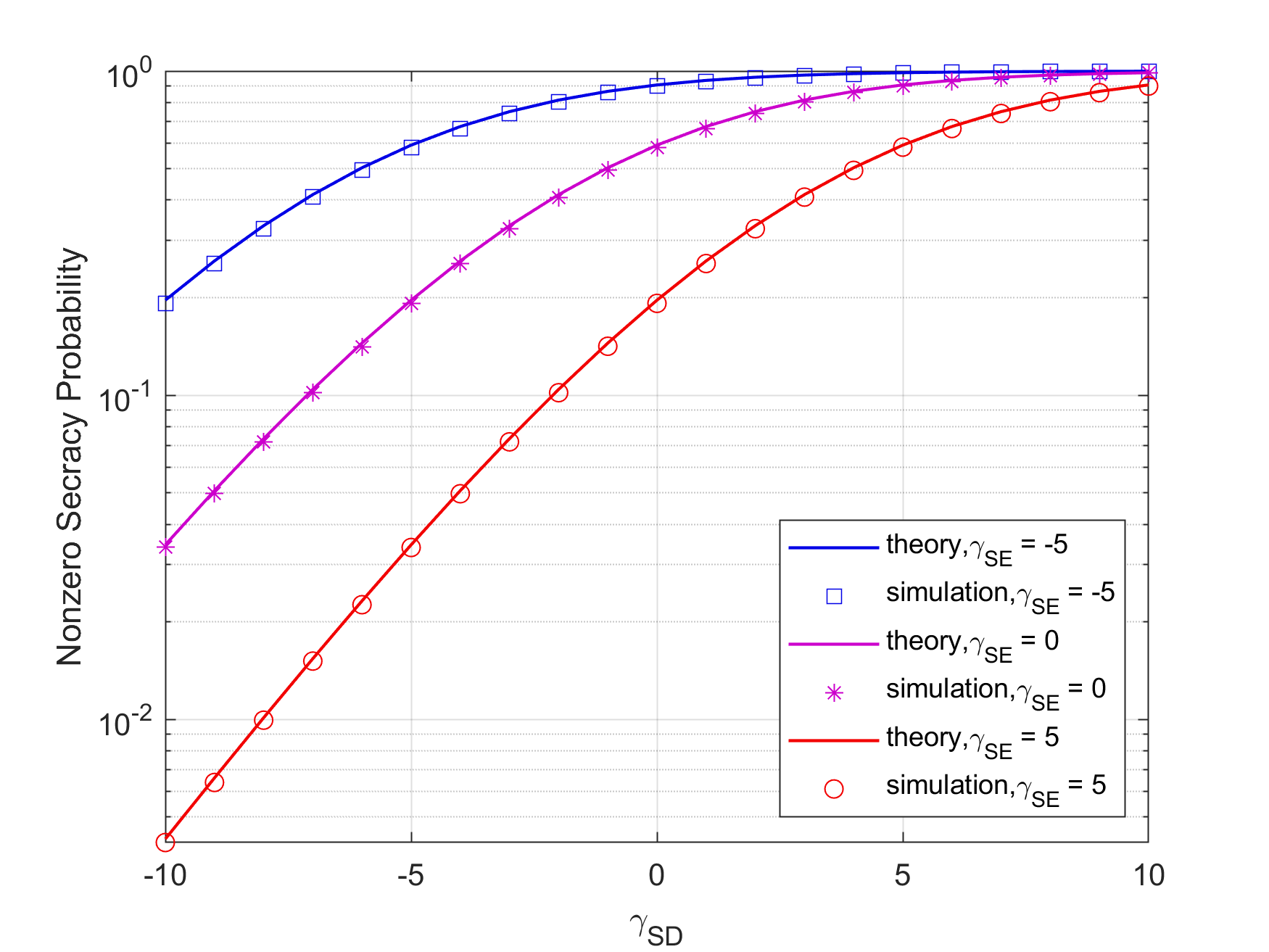} 
\caption{Non-zero secrecy probability for $\rho=o.5$ and $\bar{\gamma}_{SE}=-5,0,5 dB$}
  \label{fig:nonfloat}
\end{figure}
%%%%%%%%%%%%%%%%%%%%%%%%
%%%%%%%%%%%%%%%%%%%%%%%%
%Fig 3
\begin{figure}[H]
\centering
\includegraphics[scale = 0.5]{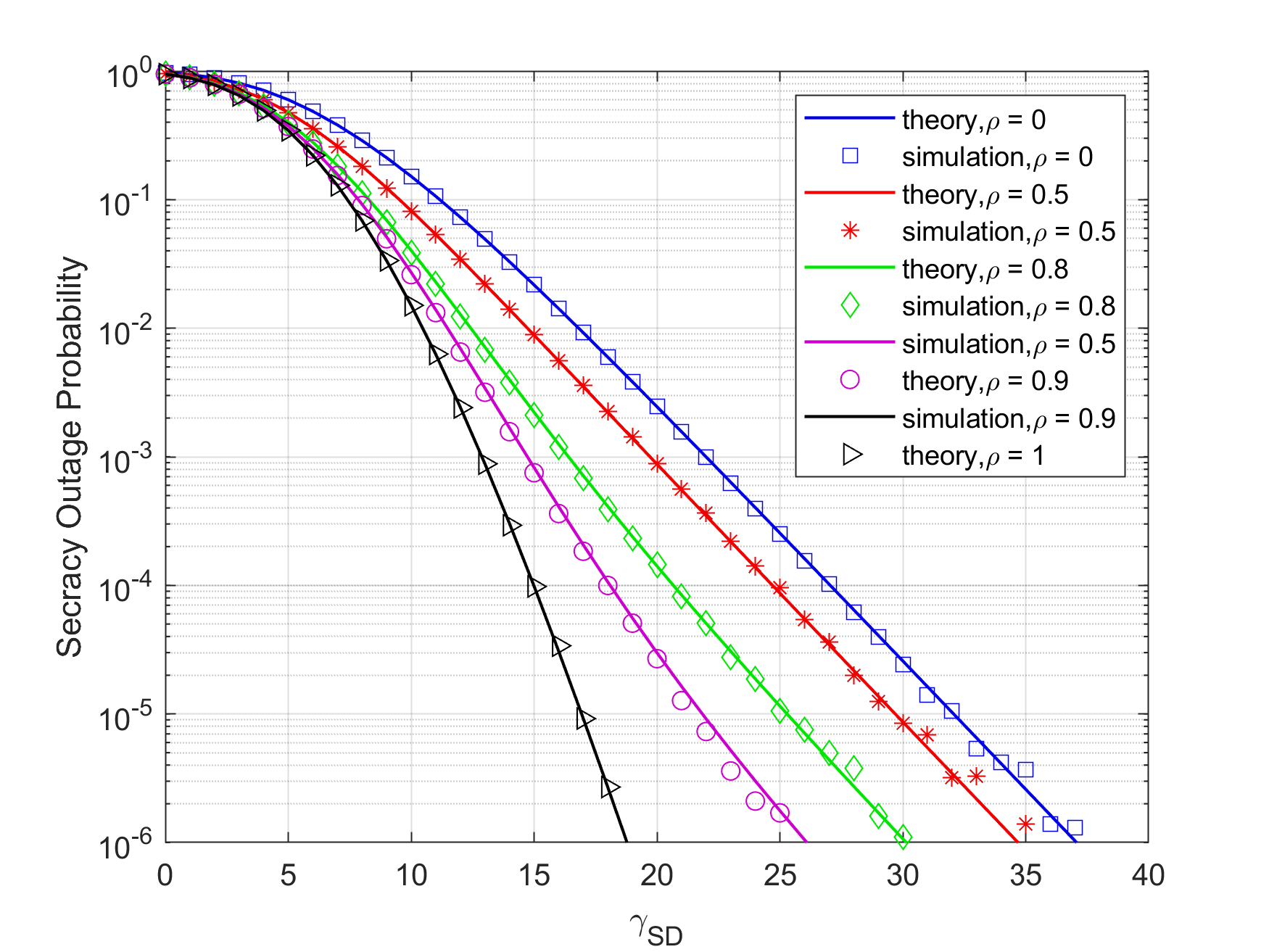} 
\caption{Secrecy Outage Probability for different values of $\rho$ and $\bar{\gamma}_{SE}=-5 dB.$ }
  \label{fig:nonfloat}
\end{figure}
%%%%%%%%%%%%%%%%%%%%%%%
%Fig 4
\begin{figure}[H]
\centering
\includegraphics[scale = 0.5]{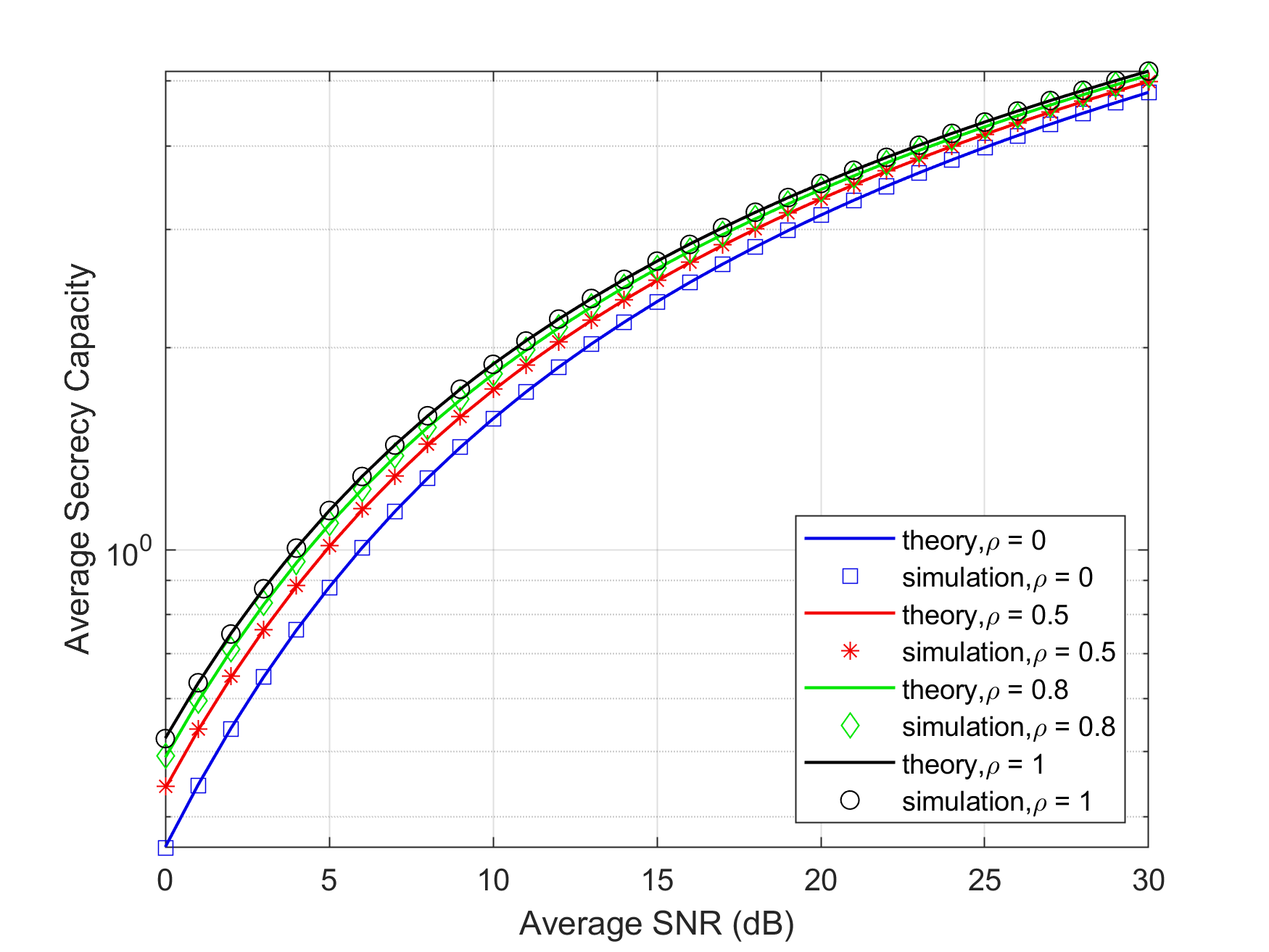} 
\caption{Average Secrecy Capacity for different values of $\rho$ and $\bar{\gamma}_{SE}=-5$.}
  \label{fig:nonfloat}
\end{figure}
%%%%%%%%%%%%%%%%%%%%%%%%
Fig. 1 shows the non-zero secrecy probability for different values of $\rho \in \{0,0.5,0.9,1\}$ while $\bar{\gamma}_{SE}=-5dB$.  It is clear that as the value of $\rho$ increases, the probability of non-zero secrecy capacity enhanced.  we can see that when $\bar{\gamma}_{SD}\gg \bar{\gamma}_{SE}$ then  $p(C_{s}>0)\approx 1$. Conversely when  $\bar{\gamma}_{SE}\gg \bar{\gamma}_{SD}$ then $p(C_{s}>0)$ decreases and will approximately close to $0$.
 In Fig. 2, the non-zero secrecy probability  for $\rho = 0.5$ and $\gamma_{SE}= \{-5,0,5\}$ dB is depicted. It is obvious that the better condition the eavesdropper channel has (i.e., higher values of $\bar{\gamma}_{SE}$), the lower probability of non-zero secrecy we get. When the main channel is much better than eavesdropper's, i.e., when $\bar{\gamma}_{SD}\gg\bar{\gamma}_{SE}$  for different values of eavesdropper's SNR, the probability of non-zero secrecy capacity is the same and  $p(C_{s}>0)\approx 1$.
Secrecy outage probability for various values of $\rho$ ishown in Fig. 3 for $\bar{\gamma}_{SE}=0$ dB. From the figure, we can see that for bigger values of $\rho$ we get lower secrecy outage probability as better CSI helps to select the best relay for the secure transmission. Also, the SOP improves with  larger values of average SNRs. Fig.4 depicts SOP for $\rho=0.5$ and $\bar{\gamma}_{SE}= -5 , 0, 5$ dB. It is clear that the outage probability increases for higher values of $\bar{\gamma}_{SE}$. 
%%%%%%%%%%%%%%%%%%%%%%%%
%Fig 5
\begin{figure}[H]
\centering
\includegraphics[scale = 0.53]{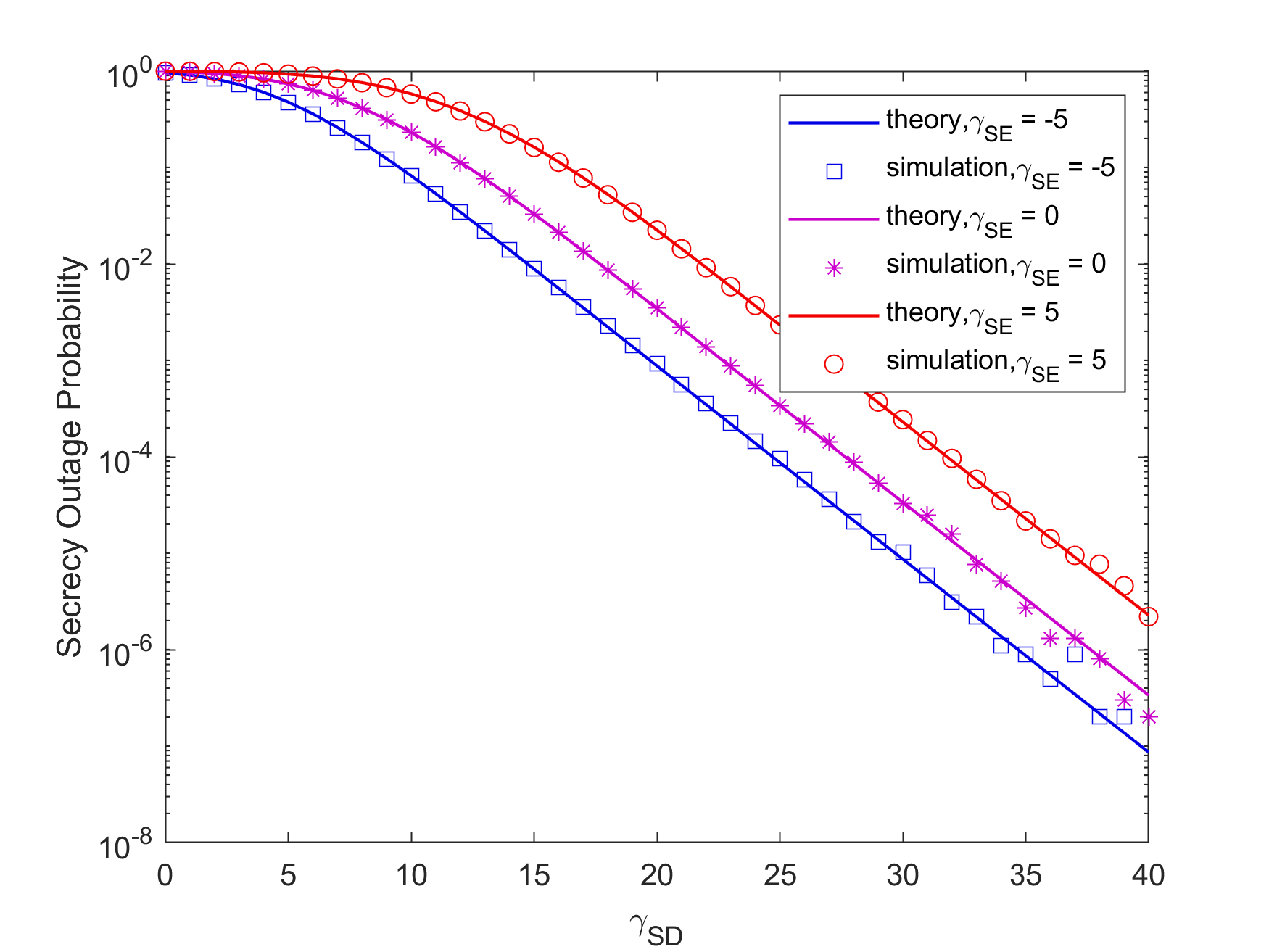} 
\caption{Secrecy capacity for different values of  $\bar{\gamma}_{SE}$ and $\rho$=0.5}
  \label{fig:nonfloat}
\end{figure}
%%%%%%%%%%%%%%%%%%%%%%%%

Fig. 5 demonstrates the average secrecy capacity for four numbers of values for  $\rho$ namely $\rho=0$, $\rho=0.5$, $\rho=0.8$, and $\rho=1$. We can see that the average secrecy capacity is increasing with the increase in values of $\rho$. In Fig. 6, the average secrecy capacity considering  $\bar{\gamma}_{SE}= -5$ dB is shown. As for non-zero secrecy capacity, for different values of $\bar{\gamma}_{SE}$, the average secrecy capacity also converges to one value as average SNR increases.

%%%%%%%%%%%%%%%%%%%%%%%
%Fig 6
\begin{figure}[H]
\centering
\includegraphics[scale = 0.53]{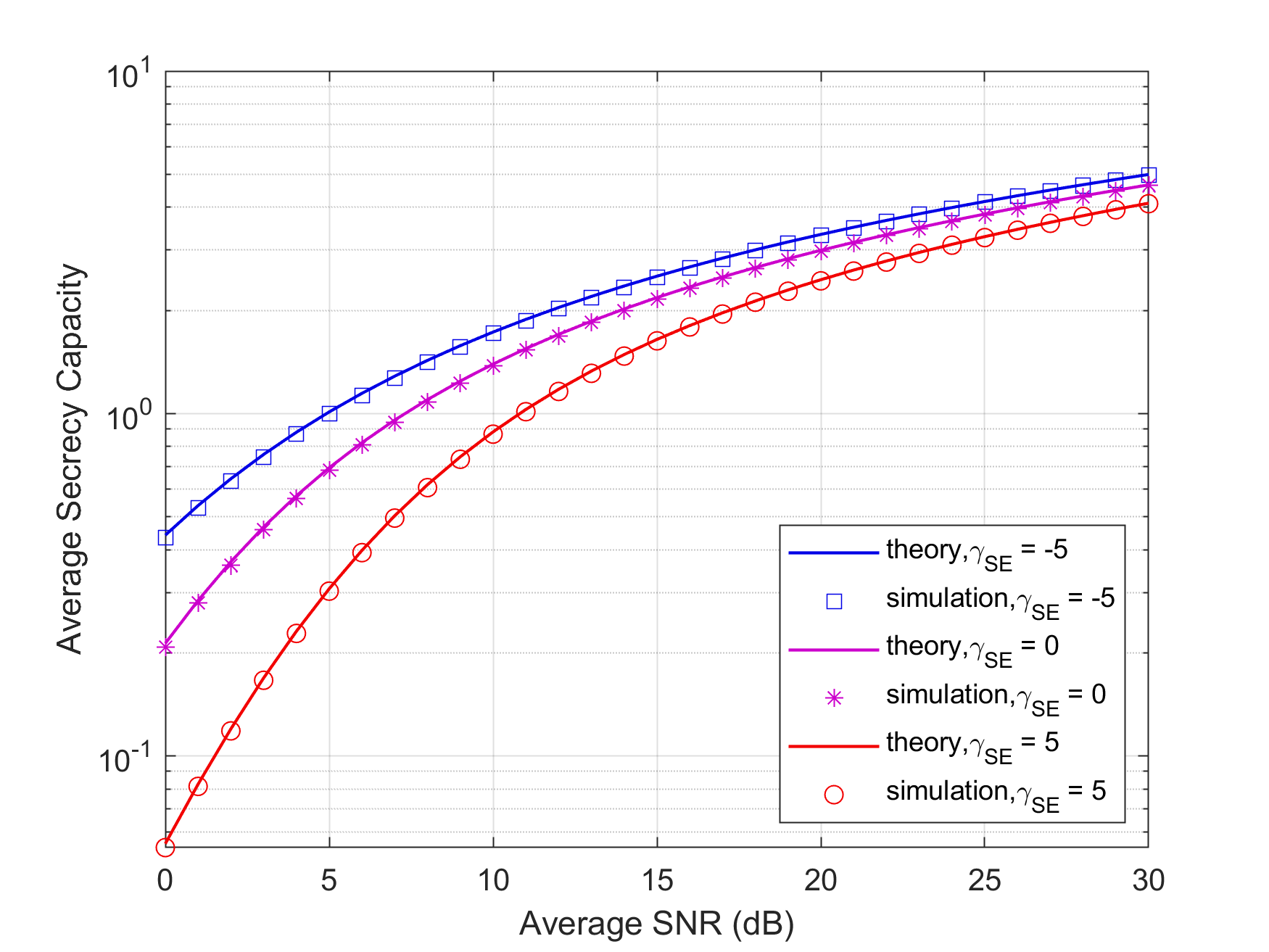} 
\caption{Average Secrecy Capacity for different values of  $\bar{\gamma}_{SE}$ and $\rho=0.5$.}
  \label{fig:nonfloat}
\end{figure}
%%%%%%%%%%%%%%%%%%%%%%%

\section{Conclusion}
 In this paper, we have investigated the secrecy performance of a relay network in the presence of an eavesdropper. The effects of outdated CSI, which is caused because of feedback delays, is considered in the opportunistic relay selection strategy. The closed-form Analytical expressions of non-zero secrecy capacity, secrecy outage probability, and average secrecy capacity are derived. The results show that for when the destination average SNR is high enough, the ergodic secrecy capacity and the non-zero secrecy capacity converge to a fixed value regardless of the value of eavesdropper's average SNR. The numerical results prove the correctness of our derived formulas.

% or


\appendices
\section{Proof of the probability of non-zero capacity}
\begin{equation*}
\begin{split}\label{eq:15}
&p(C_s>0)=p({\gamma }_{opr}>{\gamma }_{opr,e})\\
&=\int^{\infty }_0{\int^{{\gamma }_{opr}}_0{f_{{\gamma }_{opr}}({\gamma }_{opr})f_{{\gamma }_{opr,e}}({\gamma }_{opr,e}){{d{\gamma }_{opr,e}}}{{d{\gamma }_{opr}}}}}
\end{split}
\end{equation*}

\begin{align}
\begin{split}\label{eq:15}
&=\int^{\infty }_0f({\gamma }_{opr})\bigg(\frac{{\overline{\gamma }}_{ce}}{{\overline{\gamma }}_{ce}-{\overline{\gamma }}_{SE}}\Big[1-\mathrm{exp}\Big(\frac{-{\gamma }_{opr}}{{\overline{\gamma }}_{ce}}\Big)\Big]\\
&\quad\quad\quad -\frac{{\overline{\gamma }}_{SE}}{{\overline{\gamma }}_{ce}-{\overline{\gamma }}_{SE}}\Big[1-\mathrm{exp}\Big(\frac{-{\gamma }_{opr}}{{\overline{\gamma }}_{SE}}\Big)\Big]\bigg)d{\gamma }_{opr}\\
&=\sum^N_{n=0}{\left(\genfrac{}{}{0pt}{}{N}{n}\right)\frac{{\left(-1\right)}^{n-1}}{g-{\overline{\gamma }}_{SD}}}\frac{1}{{\overline{\gamma }}_{ce}-{\overline{\gamma }}_{SE}}\\
&\quad\quad\times \Bigg({\overline{\gamma }}_{ce}  \int^{\infty }_0\Big[\mathrm{exp}\Big(\frac{-{\gamma }_{opr}}{g}\Big)-\mathrm{exp}\Big(\frac{-{\gamma }_{opr}}{{\overline{\gamma }}_{SD}}\Big)\Big]\\
&\quad  \quad \quad \quad \quad\quad \quad\times\Big[1-\mathrm{exp}\Big(\frac{-{\gamma }_{opr}}{{\overline{\gamma }}_{ce}}\Big)\Big]d{\gamma }_{opr}\\
& \quad \qquad -{\overline{\gamma }}_{SE}\int^{\infty }_0\Big[\mathrm{exp}\Big(\frac{-{\gamma }_{opr}}{g}\Big)-\mathrm{exp}\Big(\frac{-{\gamma }_{opr}}{{\overline{\gamma }}_{SD}}\Big)\Big]\\
& \quad \qquad \quad \qquad \quad \times \Big[1-\mathrm{exp}\Big(\frac{-{\gamma }_{opr}}{{\overline{\gamma }}_{SE}}\Big)\Big]d{\gamma }_{opr}\Bigg)
\end{split}
\end{align}

Calculating the integrals in (\ref{eq:15}) and simplifying it, gives (\ref{eq:9}).
%\begin{align}
%\begin{split}
%&p\left(C_s>0\right)=\sum_{n=1}^N\binom{N}{n}{(-1)}^{n-1}\Big(1-\frac{1}{b-{\bar{\gamma{}}}_{SD}}\times{}\frac{1}{{\bar{\gamma{}}}_{ce}-{\bar{\gamma{}}}_{SE}}\times{}\\&\Big\{{\bar{\gamma{}}}_{ce}\times{}[\frac{1}{\frac{1}{b}+\frac{1}{{\bar{\gamma{}}}_{ce}}}-\frac{1}{\frac{1}{{\bar{\gamma{}}}_{SD}}\frac{1}{{\bar{\gamma{}}}_{ce}}}]-{\bar{\gamma{}}}_{SE}\times{}[\frac{1}{\frac{1}{b}+\frac{1}{{\bar{\gamma{}}}_{SE}}}-\frac{1}{\frac{1}{{\bar{\gamma{}}}_{SD}}+\frac{1}{{\bar{\gamma{}}}_{SE}}}]\Big\}\Big)
%\end{split}
%\end{align}

\section{Proof of the secrecy outage probability }
The SOP can be calculated as 
\begin{align}
\begin{split}\label{eq:16}
pr(C_s<R_s)&=pr\Big(C_s<R_s\vert{}{\gamma{}}_{opr}>{\gamma{}}_{opr,e}\Big)p\Big({\gamma{}}_{opr}>{\gamma{}}_{opr,e}\Big)\\
&+pr\Big(C_s<R_s\vert{}{\gamma{}}_{opr}<{\gamma{}}_{opr,e}\Big)p\Big({\gamma{}}_{opr}<{\gamma{}}_{opr,e}\Big)\\
&=pr\Big(C_s<R_s\vert{}{\gamma{}}_{opr}>{\gamma{}}_{opr,e}\Big)p\Big({\gamma{}}_{opr}>{\gamma{}}_{opr,e}\Big)\\
&+1\times{}p\Big({\gamma{}}_{opr}<{\gamma{}}_{opr,e}\Big)
\end{split}
\end{align}

Substituting (\ref{eq:5}) in (\ref{eq:16}) gives 
\begin{align}
\begin{split}
&pr\Big(C_s<R_s\vert{}{\gamma{}}_{opr}>{\gamma{}}_{opr,e}\Big)\\
&=pr\Big(\log_{2}{(1+{\gamma{}}_{opr})}-\log_{2}{(1+{\gamma{}}_{opr,e})}<R_s\vert{}{\gamma{}}_{opr}>{\gamma{}}_{opr,e}\Big)\\
&=pr\Big({\gamma{}}_{opr}<2^{R_s}(1+{\gamma{}}_{opr,e})-1\vert{}{\gamma{}}_{opr}>{\gamma{}}_{opr,e}\Big)=\\
&\int_0^{\infty{}}\int_{{\gamma{}}_{opr,e}}^{2^{R_s}(1+{\gamma{}}_{opr,e})-1}f_{{\gamma{}}_{opr}}({\gamma{}}_{opr})f_{{\gamma{}}_{opr,e}}({\gamma{}}_{opr,e})d{\gamma{}}_{opr}d{\gamma{}}_{opr,e}\\
&=p({\gamma{}}_{opr}>{\gamma{}}_{opr,e})-\\
&\int_0^{\infty{}}\int_{2^{R_s}(1+{\gamma{}}_{opr,e})-1}^{\infty{}}f_{{\gamma{}}_{opr}}({\gamma{}}_{opr})f_{{\gamma{}}_{opr,e}}({\gamma{}}_{opr,e})d{\gamma{}}_{opr}d{\gamma{}}_{opr,e}
\end{split}
\end{align}
So the SOP can be written as 
\begin{align}
\begin{split}
&pr(C_s<R_s)=1-\\
&\int_0^{\infty{}}\int_{2^{R_s}(1+{\gamma{}}_{opr,e})-1}^{\infty{}}f_{{\gamma{}}_{opr}}({\gamma{}}_{opr})f_{{\gamma{}}_{opr,e}}({\gamma{}}_{opr,e})d{\gamma{}}_{opr}d{\gamma{}}_{opr,e}
\end{split}
\end{align}
Calculating this integral results the formula in (\ref{eq:11}).

\section{Proof of the average secrecy capacity}
The first integral in (\ref{eq:13}) can be calculated as
\begin{align}
\begin{split}\label{eq:19}
&\int_0^{\infty{}}\frac{1}{2}\log_{2}(1+{\gamma{}}_{opr})f_{{\gamma{}}_{opr}}({\gamma{}}_{opr})\frac{1}{{\bar{\gamma{}}}_{ce}-{\bar{\gamma{}}}_{SE}}\Bigg({\bar{\gamma{}}}_{ce}\times{}\\
&\Big[1-\exp\Big(\frac{-{\gamma{}}_{opr}}{{\bar{\gamma{}}}_{ce}}\Big)\Big]-{\bar{\gamma{}}}_{SE}\Big[1-\exp\Big(\frac{-{\gamma{}}_{opr}}{{\bar{\gamma{}}}_{SE}}\Big)\Big]\Bigg)d{\gamma{}}_{opr}\\
&=\frac{1}{{\bar{\gamma{}}}_{ce}-{\bar{\gamma{}}}_{SE}}\sum_{n=1}^N\binom{N}{n}\frac{{(-1)}^{n-1}}{g-{\bar{\gamma{}}}_{SD}}\\
&\times \Bigg(\int_0^{\infty{}}{\bar{\gamma{}}}_{ce}\log_{2}(1+{\gamma{}}_{opr})\Big(1-\exp\Big(\frac{-{\gamma{}}_{opr}}{{\bar{\gamma{}}}_{ce}}\Big)\Big)\\
&\qquad \times{}\exp\Big(\frac{-{\gamma{}}_{opr}}{b}\Big)d{\gamma{}}_{opr}\\
&\quad -\int_0^{\infty{}}{\bar{\gamma{}}}_{ce}\times{}\log_{2}(1+{\gamma{}}_{opr})\times{}\Big(1-\exp\Big(\frac{-{\gamma{}}_{opr}}{{\bar{\gamma{}}}_{ce}}\Big)\Big)\\
&\qquad \times{}\exp\Big(\frac{-{\gamma{}}_{opr}}{{\bar{\gamma{}}}_{SD}}\Big)d{\gamma{}}_{opr}\\
&\quad+\int_0^{\infty{}}{\bar{\gamma{}}}_{SE}\log_{2}(1+{\gamma{}}_{opr})\Big(1-\exp\Big(\frac{-{\gamma{}}_{opr}}{{\bar{\gamma{}}}_{SE}}\Big)\Big)\\
&\qquad \times{}\exp\Big(\frac{-{\gamma{}}_{opr}}{b}\Big)d{\gamma{}}_{opr}\\
&\quad-\int_0^{\infty{}}{\bar{\gamma{}}}_{SE}\log_{2}(1+{\gamma{}}_{opr})\Big(1-\exp\Big(\frac{-{\gamma{}}_{opr}}{{\bar{\gamma{}}}_{SE}}\Big)\Big)\\
&\qquad \times{}\exp\Big(\frac{-{\gamma{}}_{opr}}{{\bar{\gamma{}}}_{SD}}\Big)d{\gamma{}}_{opr}\Bigg)
\end{split}
\end{align}
We define $\mathrm{Ei}(x)=\int_x^{\infty}\frac{\exp(-t)}{t}dt$. Then we have 
\begin{align}
\begin{split}\label{eq:20}
\int_0^{\infty{}}\log_{2}(1+x)\exp\Big(\frac{-x}{\alpha{}}\Big)dx=\alpha{}\log_{2}(e) \exp\Big(\frac{1}{\alpha}\Big)\mathrm{Ei}\Big(\frac{1}{\alpha{}}\Big)
\end{split}
\end{align}
Substituting (\ref{eq:20}) in (\ref{eq:19}) and calculating the integrals and similarly calculating the second part of (\ref{eq:13}), we obtain the formula in (\ref{eq:14}).


\ifCLASSOPTIONcaptionsoff
  \newpage
\fi


\begin{thebibliography}{1}

\bibitem{IEEEhowto:Pandey}
A. Pandey and S. Yadav, "Physical Layer Security in Cooperative AF Relaying Networks With Direct Links Over Mixed Rayleigh and Double-Rayleigh Fading Channels," 
\textit{IEEE Transactions on Vehicular Technology}, vol. 67, no. 11, pp. 10615-10630, Nov. 2018.

%%%%%%%%%%%%%%%%%%%

\bibitem{IEEEhowto:Feng}
Y. Feng, S. Yan, Z. Yang, N. Yang and J. Yuan, "User and Relay Selection With Artificial Noise to Enhance Physical Layer Security,"
\textit{IEEE Transactions on Vehicular Technology}, vol. 67, no. 11, pp. 10906-10920, Nov. 2018.

%%%%%%%%%%%%%%%%%%%

\bibitem{IEEEhowto:Qing}
L. Qing, H. Guangyao and F. Xiaomei, "Physical Layer Security in Multi-Hop AF Relay Network Based on Compressed Sensing," 
\textit{IEEE Communications Letters}, vol. 22, no. 9, pp. 1882-1885, Sept. 2018.

%%%%%%%%%%%%%%%%%%%%

\bibitem{IEEEhowto:Hoang}
T. M. Hoang, T. Q. Duong, N. Vo and C. Kundu, "Physical Layer Security in Cooperative Energy Harvesting Networks With a Friendly Jammer," 
\textit{IEEE Wireless Communications Letters}, vol. 6, no. 2, pp. 174-177, April 2017.

%%%%%%%%%%%%%%%%%%%

\bibitem{IEEEhowto:Shi}
H. Shi, Y. Cai, D. Chen, J. Hu, W. Yang and W. Yang, "Physical Layer Security in an Untrusted Energy Harvesting Relay Network,"
\textit{IEEE Access}, vol. 7, pp. 24819-24828, 2019.

%%%%%%%%%%%%%%%%%%%%%

\bibitem{IEEEhowto:Yadav}
S. Yadav and P. K. Upadhyay, "Impact of Outdated Channel Estimates on Opportunistic Two-Way ANC-Based Relaying With Three-Phase Transmissions,"
\textit{IEEE Transactions on Vehicular Technology}, vol. 64, no. 12, pp. 5750-5766, Dec. 2015.

%%%%%%%%%%%%%%%%%%

\bibitem{IEEEhowto:Zou}
Y. Zou, X. Wang and W. Shen, "Optimal Relay Selection for Physical-Layer Security in Cooperative Wireless Networks," 
 \textit{IEEE Journal on Selected Areas in Communications},  vol. 31, no. 10, pp. 2099-2111, October 2013.

%%%%%%%%%%%%%%%%%%

\bibitem{IEEEhowto:Huang}
K. S. Huang, ``Performance Analysis of Secrecy Outage Probability for AF-Based Partial Relay Selection with Outdated Channel Estimates,'' \textit{Mobile Information System}.

%%%%%%%%%%%%%%%%%%

\bibitem{IEEEhowto:Fan}
L. Fan, X. Lei, N. Yang, T. Q. Duong and G. K. Karagiannidis, "Secrecy Cooperative Networks With Outdated Relay Selection Over Correlated Fading Channels," 
 \textit{IEEE Transactions on Vehicular Technology}, vol. 66, no. 8, pp. 7599-7603, Aug. 2017.

%%%%%%%%%%%%%%%%%%

\bibitem{IEEEhowto:Al-Qahtani}
F. S. Al-Qahtani, C. Zhong and H. M. Alnuweiri, "Opportunistic Relay Selection for Secrecy Enhancement in Cooperative Networks,"
\textit{ IEEE Transactions on Communications}, vol. 63, no. 5, pp. 1756-1770, May 2015.
%%%%%%%%%%%%%%%%%%
\bibitem{IEEEhowto:Wang}
L. Wang, Y. Cai, Y. Zou, W. Yang and L. Hanzo, "Joint Relay and Jammer Selection Improves the Physical Layer Security in the Face of CSI Feedback Delays," 
\textit{IEEE Transactions on Vehicular Technology },vol. 65, no. 8, pp. 6259-6274, Aug. 2016.

%%%%%%%%%%%%%%%%%%

\bibitem{IEEEhowto:T1}
M. Torabi, W. Ajib and D. Haccoun, "Performance Analysis of Amplify-and-Forward Cooperative Networks with Relay Selection over Rayleigh Fading Channels," 
\textit{VTC Spring 2009 - IEEE 69th Vehicular Technology Conference }Barcelona, 2009, pp. 1-5.

%%%%%%%%%%%%%%%%%%

\bibitem{IEEEhowto:T2}
M. Torabi and D. Haccoun, "Capacity Analysis of Opportunistic Relaying in Cooperative Systems with Outdated Channel Information,"
\textit{IEEE Communications Letters}, vol. 14, no. 12, pp. 1137-1139, December 2010.

%%%%%%%%%%%%%%%%%%

\end{thebibliography}
\end{document}